\documentclass{iau} 
\usepackage{graphicx}

\title[Milky Way structure from RR Lyrae stars]
{RR Lyrae stars as probes of the Milky Way structure and formation}

\author[Pawel Pietrukowicz \& OGLE]
{Pawel Pietrukowicz$^1$ \& OGLE collaboration$^1$}

\affiliation{$^1$Warsaw University Observatory, Al. Ujazdowskie 4, 00-478 Warszawa, Poland
\\ {\tt pietruk@astrouw.edu.pl}}

\pubyear{2015}
\volume{317}
\jname{The General Assembly of Galaxy Halos: \\ Structure, Origin and Evolution}
\editors{A. Bragaglia, M. Arnaboldi, M. Rejkuba \& D. Romano, eds.}

\begin{document}

\maketitle

\begin{abstract}
RR Lyrae stars being distance indicators and tracers of old
population serve as excellent probes of the structure, formation,
and evolution of our Galaxy. Thousands of them are being discovered
in ongoing wide-field surveys. The OGLE project conducts
the Galaxy Variability Survey with the aim to detect and analyze
variable stars, in particular of RRab type, toward the Galactic
bulge and disk, covering a total area of 3000 deg$^2$. Observations
in these directions also allow detecting background halo variables
and unique studies of their properties and distribution at distances
from the Galactic Center to even 40 kpc. In this contribution,
we present the first results on the spatial distribution
of the observed RRab stars, their metallicity distribution,
the presence of multiple populations, and relations with the
old bulge. We also show the most recent results from the
analysis of RR Lyrae stars of the Sgr dwarf spheroidal galaxy,
including its center, the globular cluster M54.
\end{abstract}

\firstsection 
\section{Introduction}

RR Lyrae stars are core helium-burning giants with theoretically
estimated masses in a range from about 0.55 to 0.80~$M_{\odot}$
and ages $>$10~Gyr (\cite [Marconi et al. 2015]{Marconi_etal2015}).
These pulsating stars have spectral types from A2 to F6 or effective
temperatures between 6500 and 9000~K and $V$-band absolute magnitudes
in a range from $+0.3$ to $+0.9$~mag. RR Lyrae stars can be found
everywhere in our Galaxy. Thousands of them have been discovered
in wide-field surveys such as: ASAS, Catalina, MACHO, NSVS, OGLE,
PTF, QUEST, SDSS, SEKBO, VVV. RR Lyrae stars are divided into
fundamental-mode (type RRab), first-overtone (type RRc), and
rarely found double-mode (type RRd) pulsators. RRab stars are
on average intrinsically brighter and have higher amplitudes than
RRc stars. More importantly, RRab variables with their characteristic
saw-shaped light curves, in comparison to nearly-sinusoidal light
curves of RRc stars, are hard to overlook making the searches
for this type of variables highly complete. RRab stars have
also a very practical photometric property. Based on the pulsation
period and shape of the light curve one can estimate
metallicity of the star (\cite[Jurcsik 1995]{Jurcsik1995};
\cite[Jurcsik \& Kov\'acs 1996]{JurcsikKovacs1996};
\cite[Smolec 2005]{Smolec2005}).

The OGLE project (the Optical Gravitational Lensing Experiment)
is a long-term variability survey which started in 1992
with the original aim to detect microlensing events toward the
Galactic bulge (\cite[Udalski et al. 1992]{Udalski_etal1992}).
Since the installation of a 32-chip camera with 1.4~deg$^2$ field
of view in 2010, the project has been in its fourth phase
(OGLE-IV, \cite[Udalski et al. 2015] {Udalski_etal2015})
and focuses on large-scale monitoring. Currently, OGLE monitors
about 1.3 billion stars located in dense regions of the sky such
as the Galactic bulge, Galactic disk, and Magellanic Clouds,
by covering a total area of over 3000 deg$^2$. The survey is
conducted with the 1.3-m Warsaw telescope at Las Campanas Observatory,
Chile, administrated by the Carnegie Institution for Science.

Recently, \cite[Soszy\'nski et al. (2014)]{Soszynski_etal2014}
released a collection of 38,257 RR Lyrae stars detected
in the OGLE-IV Galactic bulge fields. An analysis of the subset
of 27,258 RRab stars has been published by
\cite[Pietrukowicz et al. (2015)]{Pietrukowicz_etal2015}.
According to their work, these metal-poor stars trace closely
the barred structure formed of intermediate-age red clump giants.
The obtained distance to the Galactic center (GC) from the RR Lyrae
stars is $R_0=8.27\pm0.01({\rm stat})\pm0.40({\rm sys})$~kpc,
in a very good agreement with the most recent estimates of $R_0$
from other methods. They show that the spatial distribution has the
shape of a triaxial ellipsoid with proportions 1:0.49(2):0.39(2)
and the major axis located in the Galactic plane and inclined
at an angle $i=20^{\circ}\pm3^{\circ}$ to the Sun-GC line of sight.
Another discovery is the presence of multiple old populations being
likely the result of mergers in the early history of the Milky Way.

\begin{figure}[htb]
\begin{center}
\includegraphics[width=3.0in]{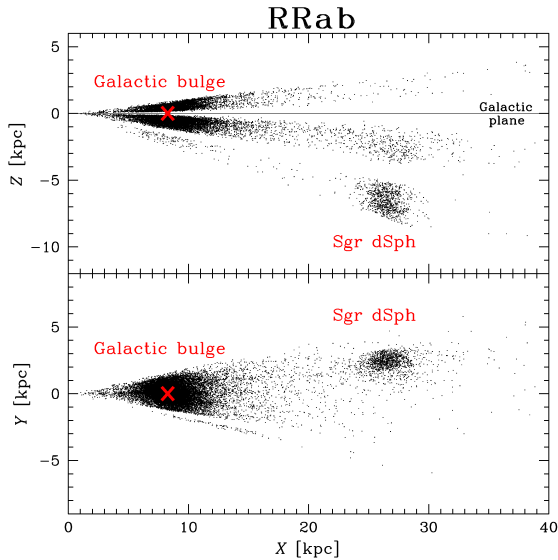}
\caption{Projection of nearly 23,000 RRab stars observed toward the OGLE-IV
bulge fields onto $XZ$ and $XY$ planes. The Sun is located at the origin
of the system. The variables are concentrated around the Galactic Center
and distributed to the outer halo. The second concentration is formed of
variables from the tidally disrupted Sgr dSph galaxy.}
\label{fig1}
\end{center}
\end{figure}

In this contribution, we report important results on far RR Lyrae
variables, namely stars observed behind the Galactic bulge.
Fig.\,\ref{fig1} shows that the distribution of stars from the
Galactic bulge area to the outer halo is smooth.
\cite[Pietrukowicz et al. (2015)]{Pietrukowicz_etal2015} has found
that the spatial density profile of bulge RR Lyrae variables can be
described as a single power law with an index of $-2.96$. This value
is very similar to indices obtained for halo stars, for example $-2.8$
for main-sequence stars near turn-off point
(\cite[Juri\'c et al. 2008]{Juric_etal2008}).

A prominent structure seen in Fig.\,\ref{fig1} at a distance about
three times larger than that to the GC is the remnant
of the Sagittarius dwarf spheroidal (Sgr dSph) galaxy.
Our search for variables within the tidal radius of M54,
the globular cluster located at the core of Sgr dSph,
has brought the detection of 277 such objects including 182
RR Lyrae stars, of which 23 are new. Based on 65 RRab stars,
very likely members of the cluster, we have estimated the distance
to M54 as $27.1\pm0.2({\rm stat})\pm1.3({\rm sys})$~kpc
(Hamanowicz et al., in prep.).

\begin{figure}[htb]
\begin{center}
\includegraphics[width=3.0in]{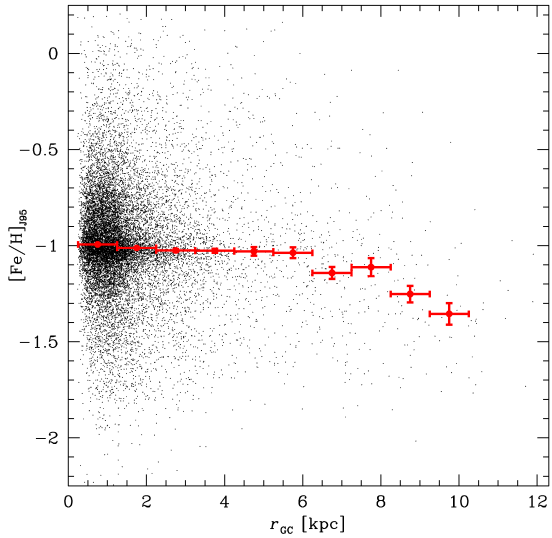}
\caption{Metallicity distribution on the \cite[Jurcsik (1995)]{Jurcsik1995}
scale as a function of distance from the Galactic center. Note a break
around 6~kpc.}
\label{fig2}
\end{center}
\end{figure}

In Fig.\,\ref{fig2}, we present radial metallicity distribution for
RRab stars from the Galactic center out to about 12~kpc. The metallicity
clearly decreases with the distance from the center but the decrease
is very mild to about 6~kpc and much more steeper farther out.
At the distance of the Sun from the GC it amounts to about $-1.1$~dex
on the \cite[Jurcsik (1995)]{Jurcsik1995} scale, in agreement
with what is observed in the solar vicinity. For instance,
a mean metallicity for 28 field RR Lyrae stars listed
in \cite[Smolec (2005)]{Smolec2005} is $-1.05$~dex.

\begin{figure}[htb]
\begin{center}
\includegraphics[angle=90,width=5.0in]{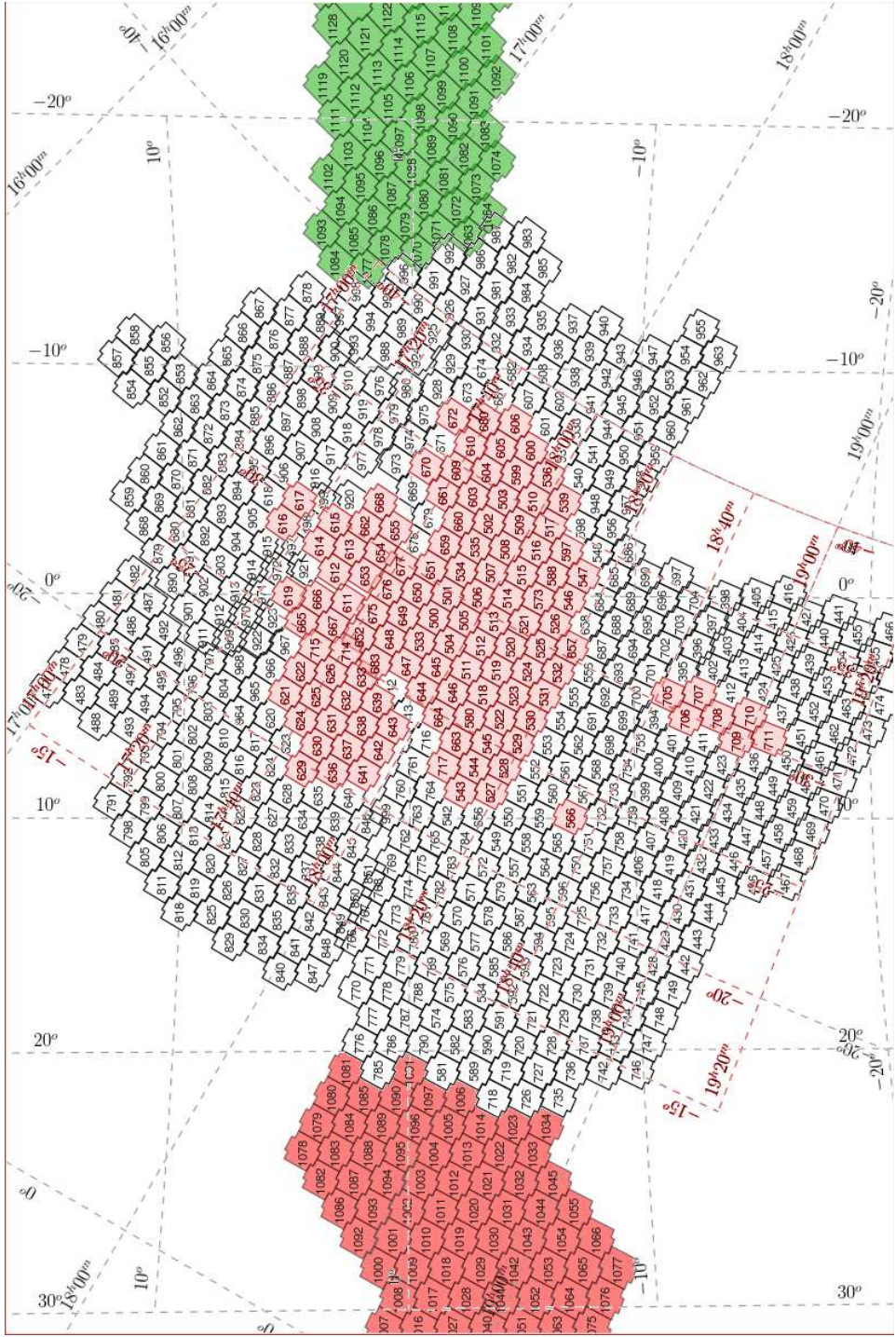}
\caption{Extended coverage of the OGLE-IV bulge area.
RR Lyrae stars from the central fields (in light pink) are analyzed
in \cite[Pietrukowicz et al. (2015)]{Pietrukowicz_etal2015}
and presented in this contribution. Fields at larger Galactic
longitudes (in green and red) represent Galactic disk fields.}
\label{fig3}
\end{center}
\end{figure}

Our results indicate that the old bulge and halo form one component
of the Galaxy. Recently, in a sample of $\sim$100 RR Lyrae stars
from the bulge area, \cite[Kunder et al. (2015)]{Kunder_etal2015} found
a high velocity object on a halo-like orbit. Ongoing and future surveys
will complete our knowledge on the shape and properties of the old
component. The OGLE survey has extended the coverage of the bulge area
with the prime aim to find and characterize RR Lyrae stars
to an angular distance of about $22^{\circ}$ from the Galactic center
(see Fig.\,\ref{fig3}).

\acknowledgments
The OGLE project has received funding from the National Science
Centre, Poland, grant MAESTRO 2014/14/A/ST9/00121 to A. Udalski.
This work has been also supported by the Polish Ministry of Sciences
and Higher Education grants No. IP2012 005672 under the Iuventus Plus
program to P. Pietrukowicz and No. IdP2012 000162 under the Ideas
Plus program to I. Soszy\'nski.

\end{document}